\documentclass{ws-ijmpd}
\usepackage{xcolor} 

\usepackage[super,compress]{cite}
\begin{document}

\renewcommand{\arraystretch}{1.3}

\markboth{G. Wang, W.-T. Ni and A.-M. Wu}
{Thruster Requirement for Various Constant Arm Space Mission Concepts}

%
\catchline{}{}{}{}{}
%

\title{Orbit design and thruster requirement for various constant-arm space mission concepts for gravitational-wave observation}

\author{Gang Wang$^\divideontimes,^\ast$, Wei-Tou Ni$^{\ast,\dagger}$ and An-Ming Wu$^\ddagger$}

\address{$^\divideontimes$Shanghai Astronomical Observatory,
Chinese Academy of Sciences, Shanghai, 200030, China \\
$^\ast$State Key Laboratory of Magnetic Resonance and Atomic and Molecular Physics, \\
Wuhan Institute of Physics and Mathematics,\\
 Innovation Academy of Precision Measurement Science and Technology (IAPM),\\ 
 Chinese Academy of Sciences, Wuhan 430071, China \\
 $^\dagger$National Astronomical Observatories, Chinese Academy of Sciences, Beijing 100012, China\\
 $^\ddagger$National Space Organization (NSPO), Science Park, Hsinchu, Taiwan, 30078, ROC \\
 gwanggw@gmail.com\\ wei-tou.ni@wipm.ac.cn \\ amwu@nspo.narl.org.tw
}

\maketitle

\begin{history}
\received{Day Month Year}
\revised{Day Month Year}
\end{history}

\begin{abstract}
In previous papers, we have addressed the issues of orbit design and thruster requirement for the constant arm versions of AMIGO (Astrodynamical Middle-frequency Interferometric Gravitational-wave Observatory) mission concept and for the constant arm GW (gravitational wave) mission concept of AIGSO (Atom Interferometric Gravitational-wave Space Observatory). In this paper, we apply similar methods to the orbit design and thruster requirement for the constant arm GW missions B-DECIGO and DECIGO, and estimate the yearly propellant requirements at the specific impulse $I_{\mathrm{sp}} = 300$ sec and $I_{\mathrm{sp}} = 1000$ sec. For the geocentric orbit options of B-DECIGO which we have explored, the fuel mass requirement is a concern. For the heliocentric orbit options of B-DECIGO and DECIGO, the fuel requirement to keep the arm equal and constant should be easily satisfied. Furthermore, we explore the thruster and propellant requirements for constant arm versions of LISA and TAIJI missions and find the fuel mass requirement is not a show stopper either. The proof mass actuation noise is a concern. To have enough dynamical range, an alternate proof mass is required. Detailed laboratory study is warranted.
\end{abstract}


\ccode{PACS numbers: 04.80.Nn, 04.80.-y, 95.10.Eg, 95.55.Ym}

\section{Introduction}

O1 and O2 observations of advanced LIGO and advanced Virgo have detected 10 GW (gravitational wave) events from stellar-size binary black hole mergers together with a binary neutron stars coalescence\cite{LVC2016,LVC2017,LVC2018}. The frequencies of these GW events are in the high-frequency band. Efforts of observations have been made in all other frequency bands – from below Hubble frequency to over terahertz frequency also\cite{Kuroda:2015owv}. 

With the successful demonstration of drag-free technology by LISA Pathfinder\cite{Armano:2016bkm,Armano:2018kix}, the technology for space GW detection is becoming mature in low-frequency and middle-frequency GW bands. In the low-frequency band, LISA is under development\cite{LISA2017}. In China, two low-frequency GW mission proposals -- TAIJI\cite{Taiji2017} and TianQin\cite{Luo2015} are under actively study. TAIJI is a mission concept in heliocentric orbit\cite{wang&ni2017}. TianQin is a mission concept in geocentric orbit\cite{Tianqin2019}. Other low or middle frequency space detection methods under conceptual study are: AIGSO (Atom Interferometric Gravitational-wave Space Observatory)\cite{Gao2017,Wang:2019oeu}, AMIGO (Astrodynamical Middle-frequency Interferometric GW Observatory)\cite{Ni:2017bzv,AMIGO}, ASTROD-GW\cite{ni2009,ni2010,ni2013,men2010a,men2010b,wang&ni2012,wang&ni2013CPB,wang&ni2015}, BBO\cite{Crowder&Cornish}, B-DECIGO\cite{B-DECIGO,Kawamura2018}, DECIGO\cite{Kawamura2018,Kawamura2006}, Super-ASTROD\cite{Ni:2008bj}, other AI (atom interferometry) proposals\cite{hogan2011,hogan2016,graham2017}, and optical clock tracking proposals\cite{loeb2015,vutha2015,Kolkowitz2016,Ebisuzaki:2018ujm}. In the middle frequency band, there are a few ground-based proposals --- MIGA\cite{chaibi2016,canuel2018}, SOGRO\cite{SOGRO,SOGRO2}, TOBA\cite{TOBA,TOBA2}, and ZAIGA\cite{Zhan:2019quq}.
  
To have significant sensitivity in the frequency band 0.1--10 Hz and yet to be a first-generation candidate for space GW missions, we have proposed a middle-frequency GW mission AMIGO\cite{Ni:2017bzv,AMIGO}. The mission concept is to use three drag-free spacecraft to form a triangular formation with nominal arm length 10,000 km, the first-generation TDI (time delay interferometry), laser power 2-10 W and telescope diameter 300-360 mm\cite{Ni:2017bzv}. The targeting sensitivity in the middle frequency band is $3 \times 10^{-21}$ Hz$^{-1}$. Four options of orbits have been studied: (i) Earth-like heliocentric orbits (3-20 degrees behind the Earth); (ii) 600,000 km high orbit formation around the Earth; (iii) 100,000 km-250,000 high orbit formation around the Earth; (iv) near Earth-Moon L4 (or L5) halo orbit formation. All four options have LISA-like formations, that is the triangular formation is nearly 60$^\circ$ inclined to the orbit plane. In 2017, we proposed this mission concept with the aim of either as a stand-alone mission or as a pathfinder with only two spacecraft. The heliocentric formation is easy to obtain; along with it, the deployment method is also obtained. When we looked deeper into orbit simulation for the geocentric orbits, we found that it was difficult to find a hoped triangular formation\cite{AMIGO}. Since the arm length is small, we tried to use thruster to keep the arms at constant equal length, and found that the fuel requirement make the geocentric AMIGO orbits technologically not feasible currently, but perfectly feasible for the heliocentric orbits. Therefore, for the heliocentric mission orbit choice of AMIGO, there can be both geodesic implementation and constant arm implementation of orbit configuration. In the AMIGO constant arm implementation for heliocentric orbit, the acceleration to maintain the formation can be designed to be less than 15 nm/s$^2$ and the thruster requirement can be smaller than 15 $\mu$N\cite{AMIGO}.

AIGSO is a mission concept using atom interferometry to detect the GWs mainly in the middle frequency band (0.1--10 Hz)\cite{Gao2017}. AIGSO proposes to have three spacecraft in linear formation with 10 km baseline. The three spacecraft maintain 5 km + 5 km constant arm-length formation. In a previous paper, we have addressed the issue of orbit design and thruster requirement for the constant arm AIGSO mission concept. The acceleration to maintain the formation can be designed to be less than 30 pm/s$^2$ and the required amplitude of thruster force will be smaller than 30 nN\cite{Wang:2019oeu}.

B-DECIGO of arm length 100 km and DECIGO of arm length 1000 km are constant arm GW missions due to its Fabry-Perot implementation. Since arm length of B-DECIGO is shorter than AMIGO by 2 orders of magnitude, we consider three shrunk orbit configurations of AMIGO as the possible orbits of B-DECIGO. By using these downscaled orbits, we estimate the propellant requirement to maintain the constant arm-length for B-DECIGO mission concept.

Our paper is organized as follows. In Section \ref{sec:2}, we briefly describe the algorithm to attain the selected orbits. In Section \ref{sec:3}, we expatiate the method to obtain the trajectories of the S/Cs to maintain the constant arm interferometry, and to estimate the acceleration requirement for the thrusters. 
In Section \ref{subsec:B-DECIGO}, we obtain the thruster acceleration requirement for the B-DEICGO at different orbital configurations.
In Section \ref{subsec:DECIGO-result}, we find that the acceleration to maintain the formation can be designed to be less than 0.2 nm/s$^2$ and the thruster requirement be less than 0.2 nN for DECIGO. As far as these requirements are concerned, DECIGO is perfect feasible.
In Section \ref{subsec:NewLISA_TAIJI}, we explore a new scheme of formation control for LISA and TAIJI. In this scheme of control, a constant-arm equilateral triangle is formed by one geodesic spacecraft followed by two non-geodesic spacecraft instead of three geodesic motion spacecraft forming a nearly equilateral triangle.
In Section \ref{sec:Fuel_result}, we obtain the propellant requirement for various missions at specific impulse 300 sec and 1000 sec, and compile the results in Table \ref{Tab:results_table}.
We present feasibility discussions and deliberate on the implementation requirements in Section \ref{sec:conclusion}.

\section{Mission Orbit Selection} \label{sec:2}

\subsection{Orbit Selection} \label{sec:geodesic}
In this subsection, we briefly state the initial orbit selection algorithm used in our previous works \refcite{wang&ni2017,dnw2013,wang&ni2013CQG}, which is specifically discussed in \refcite{Dhurandhar+etal+2005}. For the B-DECIGO configurations, as we will describe in the following section, their orbits are downscaled from the selected AMIGO configurations. The initial orbit configuration of the DECIGO used in this work are generated from the Eqs. \eqref{equ:1}-\eqref{equ:initialcondition} below without any optimization.
In subsection \ref{subsec:NewLISA_TAIJI}, The initial LISA orbit configuration is picked from Ref. \refcite{wang&ni2017}, while the orbit of TAIJI is recalculated and re-optimized according to the method of Ref. \refcite{wang&ni2017} for the configuration ahead of Earth by $20^\circ$.

For a LISA-like orbit configuration which has a $60^\circ$ inclination with respect to the ecliptic plane, the orbit of each S/C has an eccentricity $e$ and inclination $\iota$. The first order parameter is $\alpha\ [ = \iota = l/(2R)]$, where $l$ is the nominal arm length and $R$ is the orbital radius. A set of S/C initial conditions in the heliocentric elliptical coordinate is defined in \refcite{Dhurandhar+etal+2005},
 \begin{equation} \label{equ:1}
 \begin{split}
   X_k & = R ( \cos \psi_k + e) \cos \epsilon \\
   Y_k & = R \sqrt{ 1 - e^2} \sin \psi_k \qquad (k = 1, 2, 3) \\
   Z_k & = R( \cos \psi_k + \epsilon ) \sin \epsilon
 \end{split}
\end{equation}
where $\epsilon \simeq 3.34 \times 10^{-9} \times l/ $km; orbital eccentricity $e \simeq 1.93 \times 10^{-9} \times l$/km;  
$R = 1$ AU; and $\psi_k$ is the eccentric anomaly which is related to the mean anomaly $ \Omega (t - t_0)$ by equation
 \begin{equation}
   \psi_k + e \sin \psi_k = \Omega (t - t_0) - (k-1) \frac{2\pi}{3},
 \end{equation}
where $\Omega$ is 2$\pi$/(one sidereal year). The $x_k , y_k, z_k (k=1,2,3)$ is defined as
 \begin{equation}
 \begin{split}
   x_k & = X_k \cos \left[ \frac{2\pi}{3} (k-1) + \varphi_0  \right] - Y_k \sin \left[\frac{2\pi}{3} (k-1) + \varphi_0 \right],\\
   y_k & = X_k \cos \left[ \frac{2\pi}{3} (k-1) + \varphi_0 \right] + Y_k \sin \left[\frac{2\pi}{3} (k-1) + \varphi_0 \right], \\
   z_k & = Z_k,
 \end{split}
\end{equation}
where $\varphi_0 = \psi_E - \theta$ and $\psi_E$ is the position angle of Earth with respect to the X-axis at initial time in the ecliptic plane.
The initial positions of the S/Cs in the heliocentric coordinate are
\begin{equation} \label{equ:initialcondition}
 \begin{split}
  \mathbf{r}_{\mathrm{S/Ck}} &= [x_k , y_k , z_k] \quad (k = 1, 2, 3) \\
 \end{split}
\end{equation}

\subsection{Vector Defination in the Instantaneous Plane} \label{sec:fixed-baseline_1}

Follow the procedures in Section \ref{sec:geodesic}, we can obtain the geodesic orbits of the three S/Cs. To identify the instantaneous plane formed by the three S/Cs, the direction $\mathbf{n}_z$ of the plane can be defined from the instantaneous positions of the S/Cs by
\begin{equation} \label{eq:vector_z}
\begin{split}
\mathbf{n}_{23} (t) &= \frac{ \mathbf{r}_{\mathrm{S/C}3} -  \mathbf{r}_{\mathrm{S/C}2}}{ | \mathbf{r}_{\mathrm{S/C}3} -  \mathbf{r}_{\mathrm{S/C}2} | }, \\
\mathbf{n}_{21} (t) &= \frac{ \mathbf{r}_{\mathrm{S/C}1} -  \mathbf{r}_{\mathrm{S/C}2}}{ | \mathbf{r}_{\mathrm{S/C}1} -  \mathbf{r}_{\mathrm{S/C}2} | }, \\
\mathbf{n}_z (t) &= \frac{ \mathbf{n}_{23} \times \mathbf{n}_{21}  }{ | \mathbf{n}_{23} \times \mathbf{n}_{21} |},
 \end{split}
\end{equation}
where $\mathbf{r}_{\mathrm{S/C}i}$ is the instantaneous positions of the S/C$i$ ($i$=1,2,3) at $t$.
\begin{figure}[htb]
   \centering
   \includegraphics[width=0.49\textwidth]{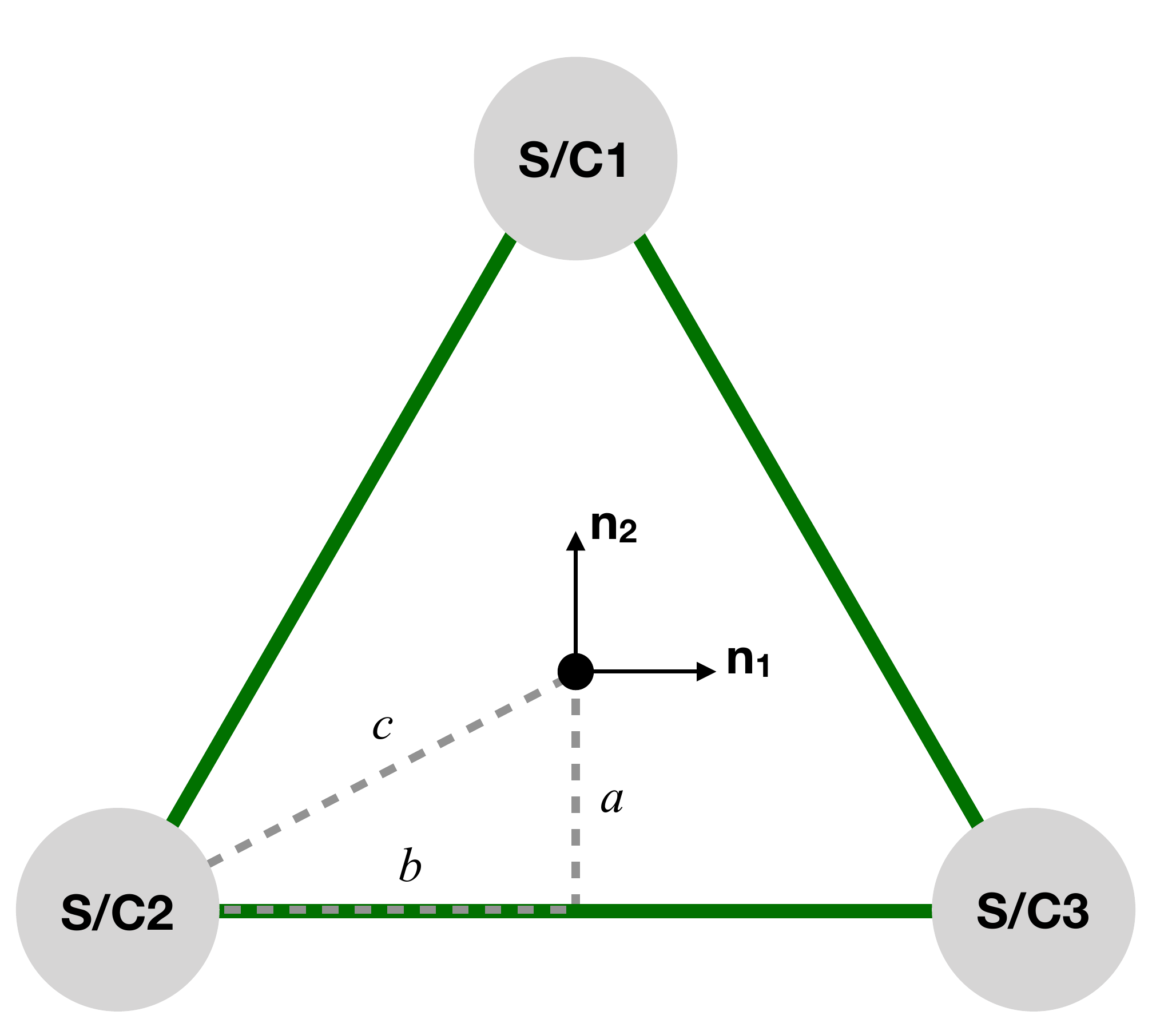}
   \caption{Diagram of the three S/Cs for a constant arm length triangle configuration.}  \label{fig:diagram}
\end{figure}

From the vectors obtained in Eq. \eqref{eq:vector_z}, we define the two orthogonal unit vectors in a instantaneous plane as shown in Fig \ref{fig:diagram},
\begin{equation} \label{equ:n_vector}
\begin{split}
\mathbf{n}_1 &=  \mathbf{n}_{23}, \\
 \mathbf{n}_2 &=  \mathbf{n}_z \times \mathbf{n}_1.
 \end{split}
\end{equation}

\section{Trajectory Choices and Thruster Calculation Algorithm} \label{sec:3}

\subsection{Trajectory Choices}
From the two vectors in Eq. \eqref{equ:n_vector}, we calculate the trajectories of the S/Cs which keep the constant equal-arm configuration as follows: one S/C follows its geodesic orbit as selected in Section \ref{sec:geodesic}, and other two S/Cs follow this S/C to form an equilateral triangle with desired constant arm length by using the thruster compensation. 

Without loss of generality, here we let the S/C2 and S/C3 follow the geodesic motion of S/C1. The trajectories of the three S/Cs are calculated by
\begin{equation} \label{equ:traj_position1}
\begin{split}
  \mathbf{r}_{\mathrm{traj, S/C1}} &= \mathbf{r}_{\mathrm{S/C}1}, \\
  \mathbf{r}_{\mathrm{traj, S/C2}} &= \mathbf{r}_{\mathrm{S/C}1} - b \mathbf{n}_1 - (a+c) \mathbf{n}_2, \\
  \mathbf{r}_{\mathrm{traj, S/C3}} &= \mathbf{r}_{\mathrm{S/C}1} + b \mathbf{n}_1 - (a+c) \mathbf{n}_2,
 \end{split}
\end{equation}
where $a = {l}/{(2\sqrt{3})}, b = l/2, c = l/\sqrt{3} $, the $l$ is the proposed arm length. $\mathbf{r}_{{\mathrm{S/C}}i}$ is the geodesic orbit of S/C$i$. Therefore, the three trajectories, $\mathbf{r}_{\mathrm{traj, S/C}i} \ (i=1,2,3)$ will form a constant arm equilateral triangle.

For the DECIGO configuration, there is cluster b (30-degree-b/60-degree-b) as shown in Fig. \ref{fig:DECIGO_orbit}. We assume the three spacecraft of cluster b will follow the geodesic S/C1 in cluster a. The trajectories of the three S/Cs are calculated by
\begin{equation} \label{equ:traj_position2b}
\begin{split}
 \mathbf{r}_{\mathrm{traj, S/C1b}} &=  \mathbf{r}_{\mathrm{S/C}1} -  2 c \mathbf{n}_2,  \\
 \mathbf{r}_{\mathrm{traj, S/C2b}} &=  \mathbf{r}_{\mathrm{S/C}1} -  b \mathbf{n}_1 + (a-c) \mathbf{n}_2, \\
 \mathbf{r}_{\mathrm{traj, S/C3b}} &=  \mathbf{r}_{\mathrm{S/C}1} + b \mathbf{n}_1 + (a-c) \mathbf{n}_2.
 \end{split}
\end{equation}

\subsection{Thruster Acceleration Algorithm}

From Eq. \eqref{equ:traj_position1} or \eqref{equ:traj_position2b}, we can obtain the acceleration at a specific point in a trajectory by calculating the second derivative of position with respect to time,
\begin{equation} \label{equ:traj_acceleration}
 \mathbf{a}_{\mathrm{traj}} = \mathbf{\ddot{r}}_{\mathrm{traj}}.
\end{equation}
On the other hand, for each trajectory, we put the position and its first derivative, velocity, into the ephemeris framework to calculate the acceleration $ \mathbf{a}_\mathrm{eph}$,
\begin{equation} \label{equ:a_eph}
  \mathbf{a}_\mathrm{eph} ( \mathbf{r}_{\mathrm{traj}} , \mathbf{\dot{r}}_{\mathrm{traj}})= \mathbf{a}_{\mathrm{Newton}} + \mathbf{a}_{\mathrm{1PN}} +  \mathbf{a}_{ \mathrm{fig}} +  \mathbf{a}_{\mathrm{asteroid}}, 
\end{equation}
where $\mathbf{a}_{\mathrm{Newton}} $ and $ \mathbf{a}_{\mathrm{1PN}} $ are the Newtonian and first-order post-Newtonian acceleration from the major celestial bodies in the solar system considered as point mass, $\mathbf{a}_{ \mathrm{fig}} $ is the acceleration due to the figure effects from the Sun, Earth and Moon, and $ \mathbf{a}_{\mathrm{asteroid}}$ is the acceleration from the 340 asteroids' Newtonian perturbation.
The explicit interactions in our CGC ephemeris framework are fully described in reference \refcite{wang&ni2012,wang&ni2015,wang2011}.

Then the thruster acceleration to maintain the constant arm length trajectories is calculated by
\begin{equation}
 \mathbf{a}_{\mathrm{thruster}} =  \mathbf{a}_{\mathrm{traj}} - \mathbf{a}_\mathrm{eph}.
\end{equation}

\section{Thruster Requirement for Different Missions} \label{sec:missions-acc-result}

\subsection{B-DECIGO} \label{subsec:B-DECIGO}

The B-DECIGO proposed constant 100 km arm length mission orbit around the Earth\cite{B-DECIGO,Kawamura2018}. In our study about AMIGO mission which proposed the arm length with 10000 km arm length, we considered a multitude of possible orbital configurations\cite{AMIGO}. In this work, the configurations AMIGO-E1, AMIGO-EML4 and AMIGO-S are selected to downscale by 100 times as the possible B-DECIGO orbital configurations. Therefore, we choose one geodesic orbit from the AMIGO configuration as fiducial, and find the other two S/Cs trajectories by scaling down the other two AMIGO arms in the constant-equal-arm formation. In this work, three B-DECIGO orbital configurations considered are as follow,
\begin{itemize}
\item B-DECIGO-AM-E1 around the Earth with semimajor axis of 100,000 km, this orbit configuration is downscaled from the AMIGO-E1 configuration\cite{AMIGO}. 
\item B-DECIGO-AM-EML4 near the Earth-Moon L4 point, this orbit configuration is shrunk from the AMIGO-EML4 configuration\cite{AMIGO}
\item B-DECIGO-AM-S which is LISA-like around the Sun with 10-degree trailing angle. In this case, the 100 km is downscaled from the AMIGO-S 10,000 km configuration\cite{AMIGO}.
\end{itemize}

\begin{figure}[htb]
   \centering
   \includegraphics[width=0.45\textwidth]{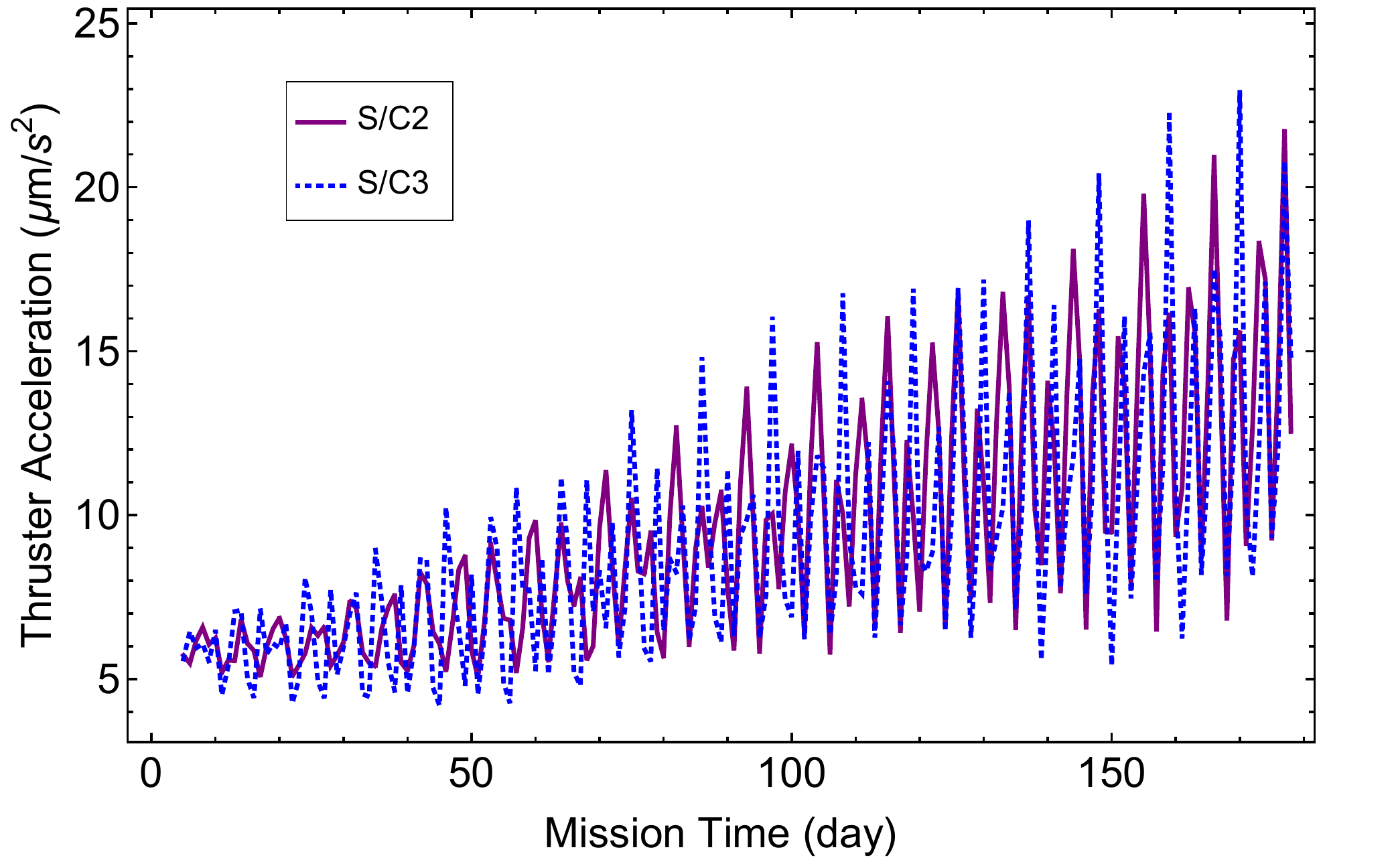} \ 
   \includegraphics[width=0.45\textwidth]{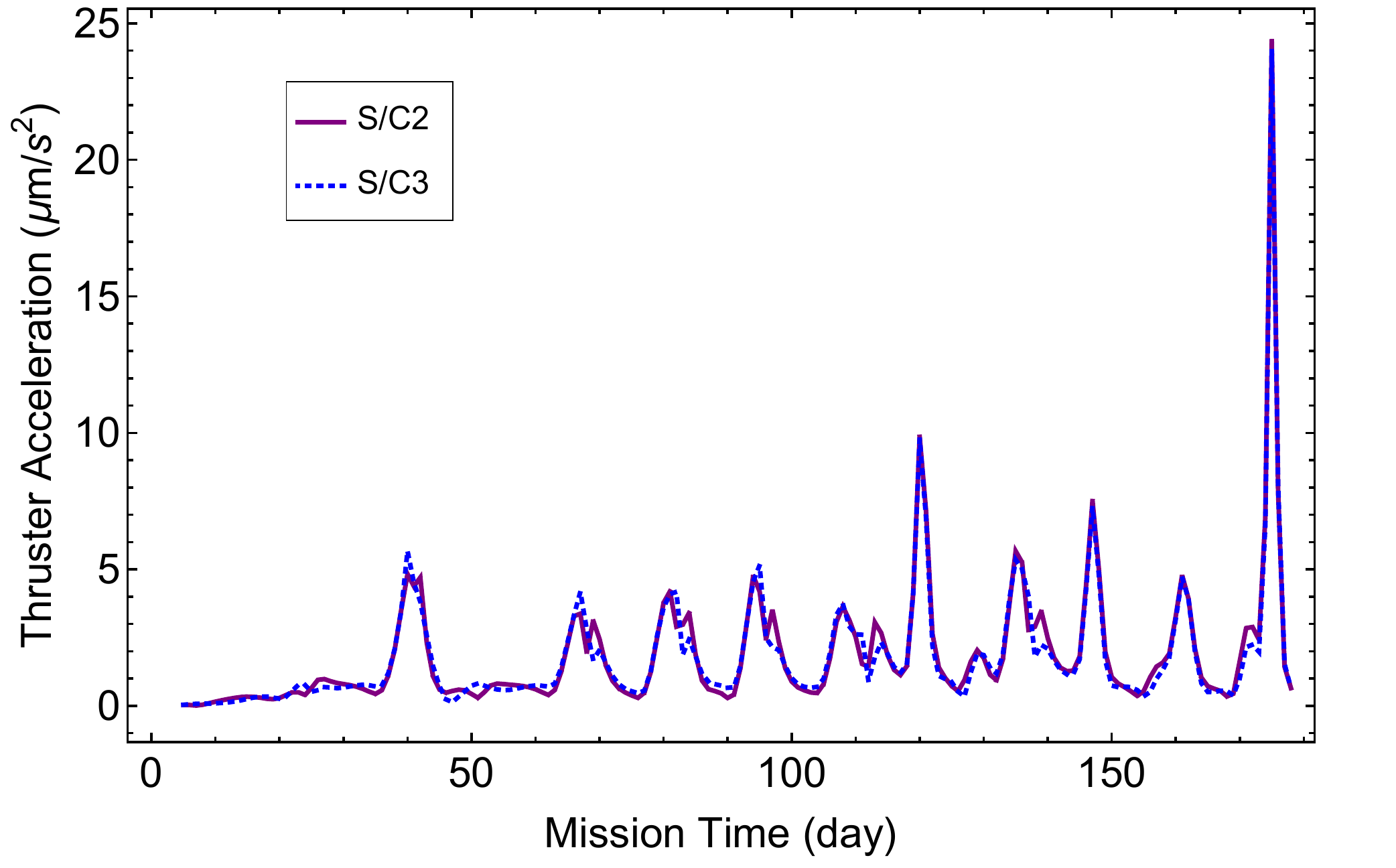} 
   \includegraphics[width=0.45\textwidth]{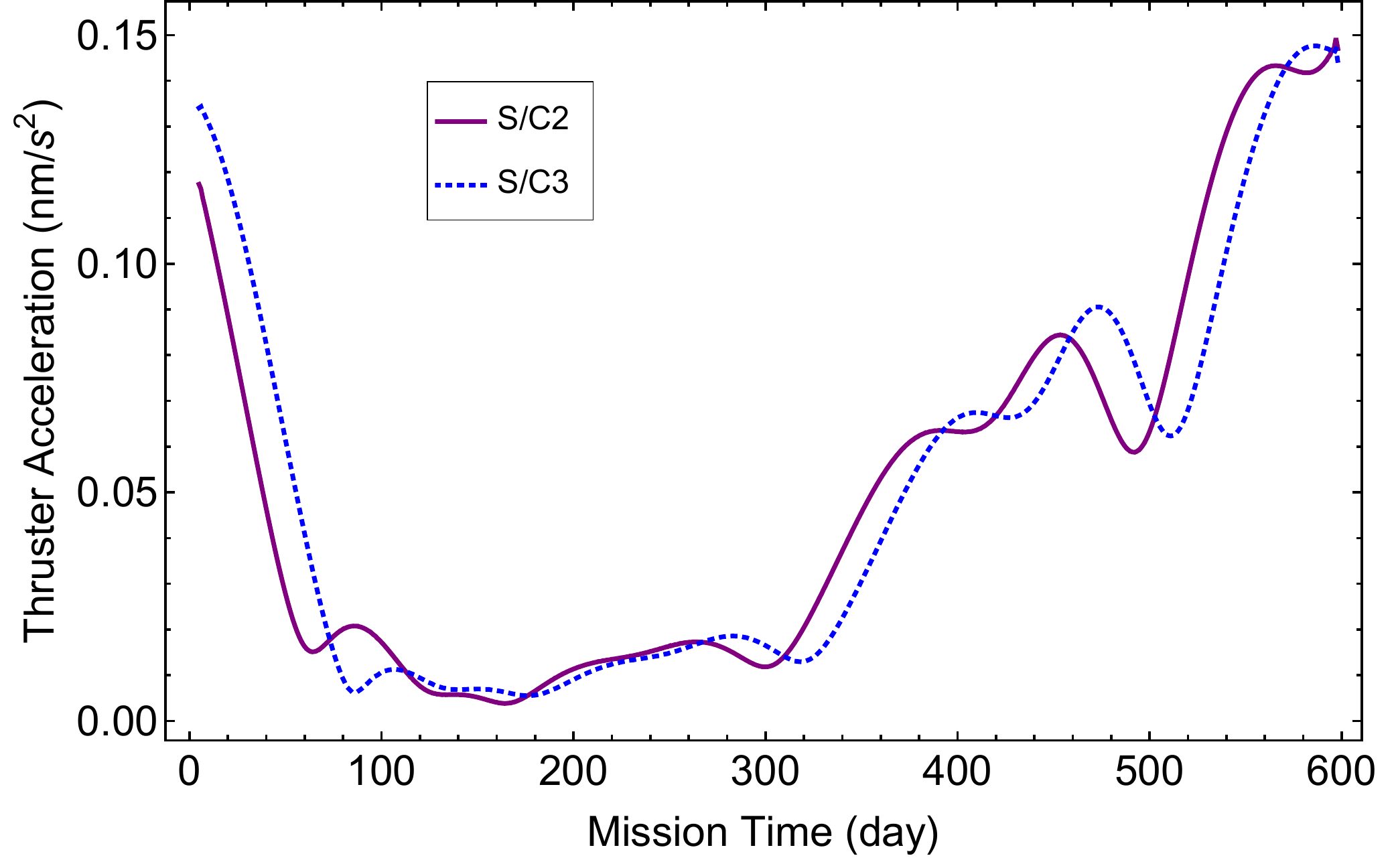} 
   \caption{The thruster acceleration compensations to maintain the 100 km arm length for S/Cs in the CGC3.0 ephemeris framework for B-DECIGO-AM-E1 (upper left panel), B-DECIGO-AM-EML4 (upper right panel) and B-DECIGO-AM-S (lower panel) configurations, respectively.}
   \label{fig:diffA_B_DECIGO}
\end{figure}

The corresponding results are shown in Fig. \ref{fig:diffA_B_DECIGO}. As we can see, the acceleration of the B-DECIGO-AM-E1 configuration is up to 25 $\mu$m/s$^2$ in 180 days, the acceleration of the B-DECIGO-AM-EML1 configuration also could go up to 25 $\mu$m/s$^2$ in 180 days, and the acceleration of the B-DECIGO-AM-S configuration is up to 0.15 nm/s$^2$ in 600 days. The results clearly show that the heliocentric orbit B-DECIGO is much easier (4 orders of magnitude) to adjust as expected.

\subsection{DECIGO} \label{subsec:DECIGO-result}

\begin{figure}[htb]
   \centering
   \includegraphics[width=0.6\textwidth]{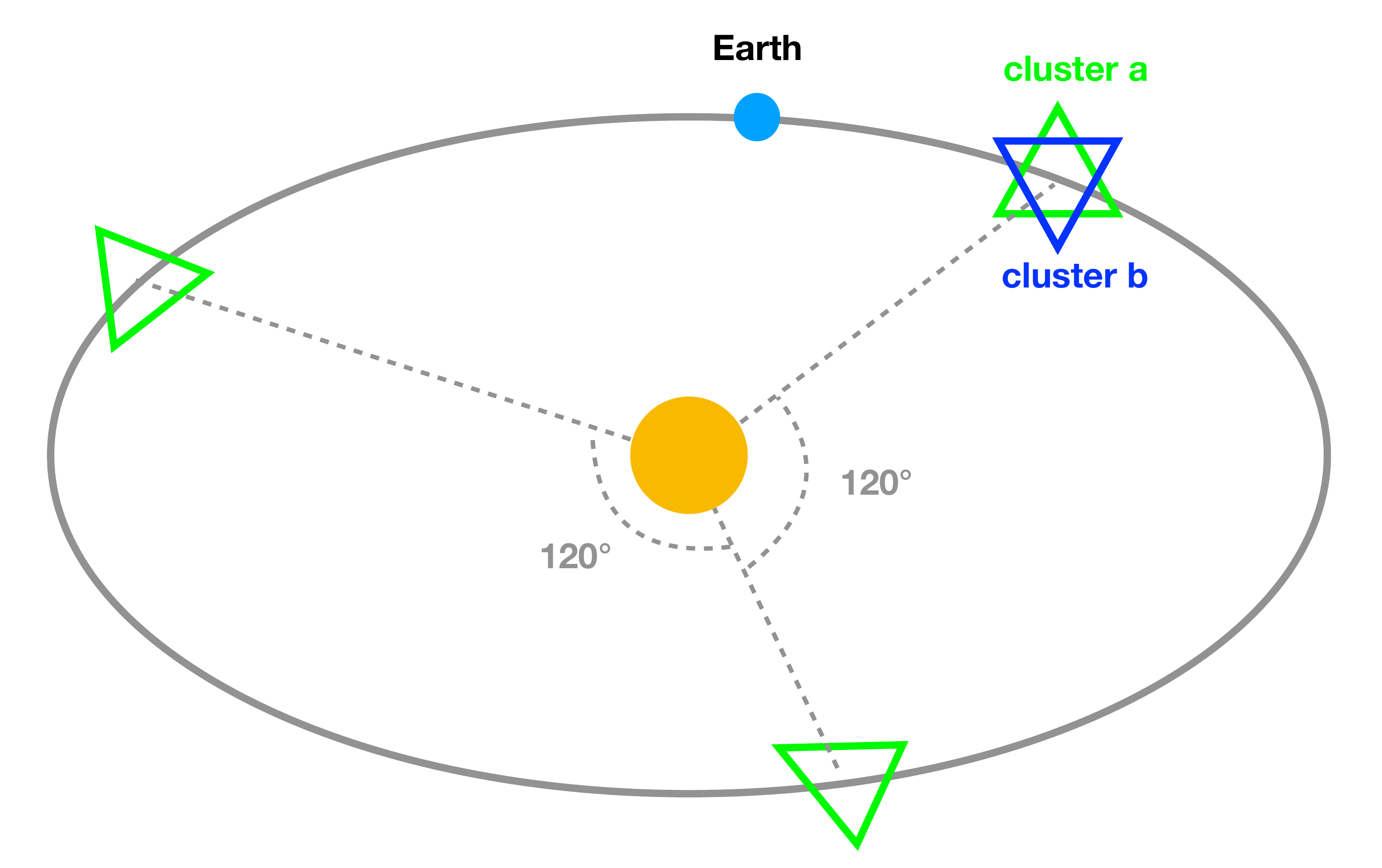} \ 
   \caption{The DECIGO orbit formation.}
   \label{fig:DECIGO_orbit}
\end{figure}

The DECIGO proposed constant 1000 km arm length in the heliocentric Earth trailing LISA-like orbit\cite{Kawamura2018}. As the diagram shown in Fig. \ref{fig:DECIGO_orbit}, six S/Cs form two concentric equilateral triangles at the same trailing angle position, and the orientation of two triangle have 180$^\circ$ difference in the formed plane. With respect to this position, there are another two triangle constellations with angle separation 120$^\circ$ and 240$^\circ$ in the ecliptic plane, respectively.

\begin{figure}[htb]
   \centering
   \includegraphics[width=0.49\textwidth]{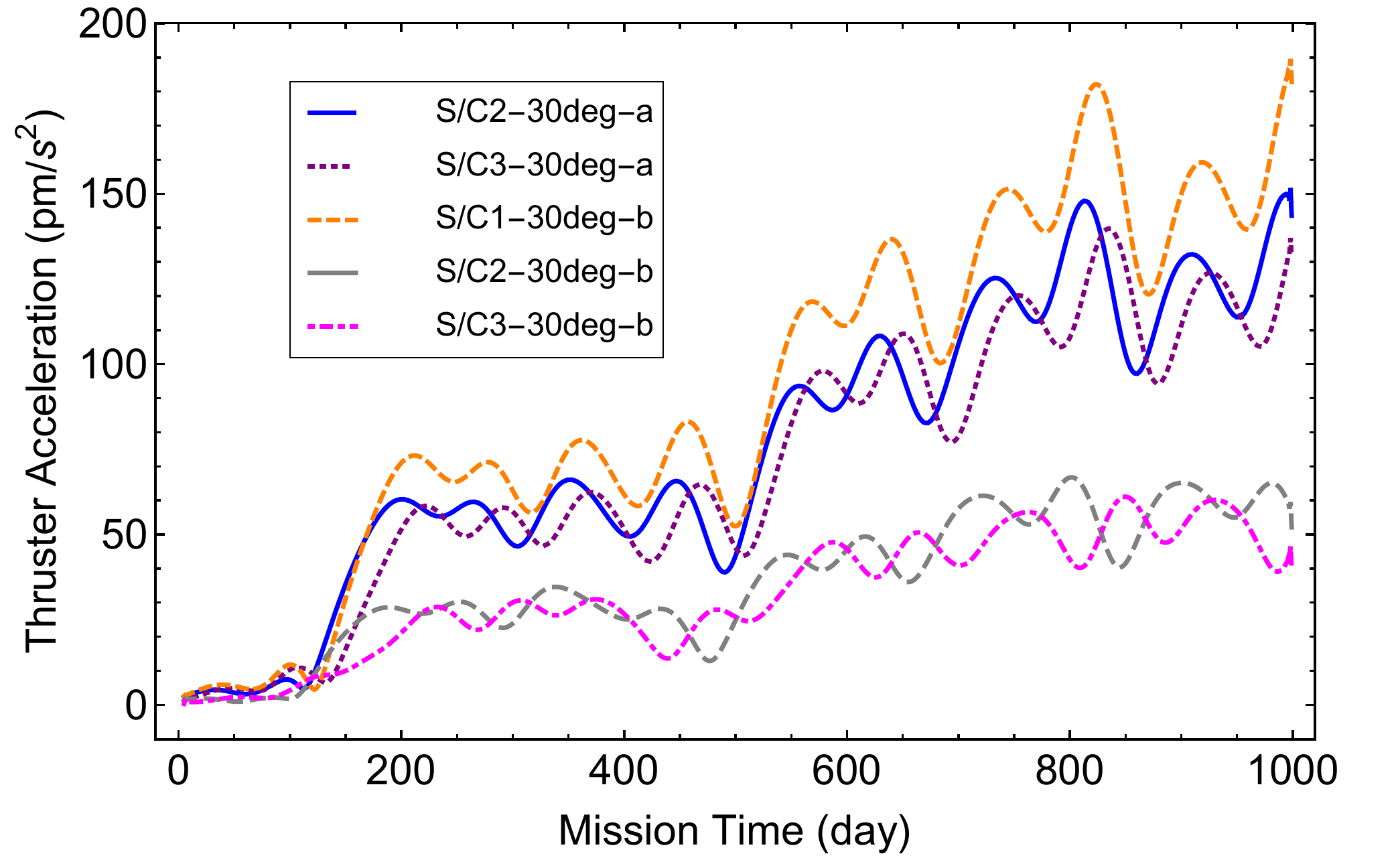} \ 
   \includegraphics[width=0.48\textwidth]{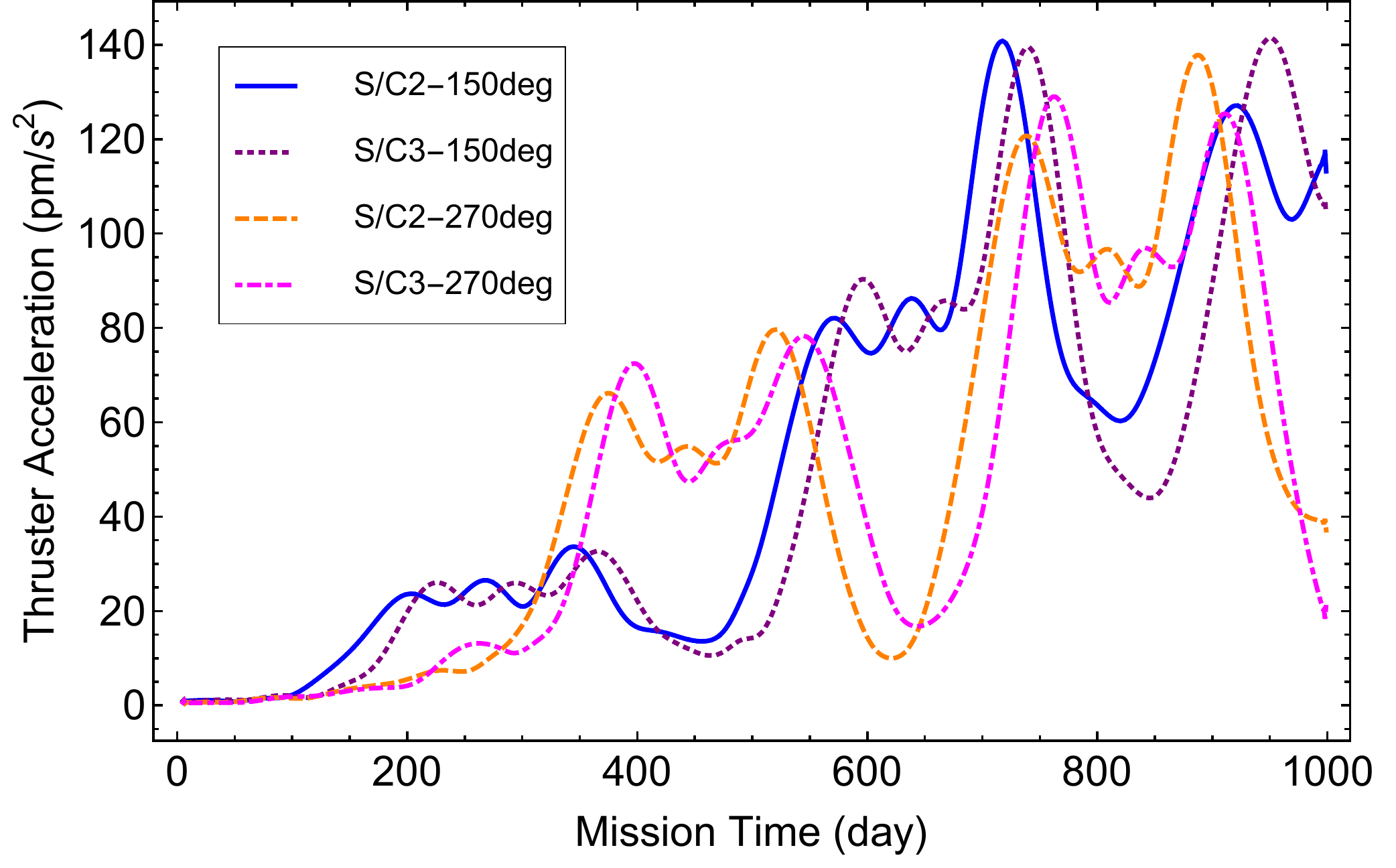}
   \caption{The thruster acceleration compensations to maintain the DECIGO arm length for S/Cs in the CGC3.0 ephemeris framework for 30-degree clusters a and b (left panel), 150-degree and 270-degree (right panel) trailing angle configurations, respectively.}
   \label{fig:diffA_DECIGO_1}
\end{figure}

\begin{figure}[htb]
   \centering
   \includegraphics[width=0.49\textwidth]{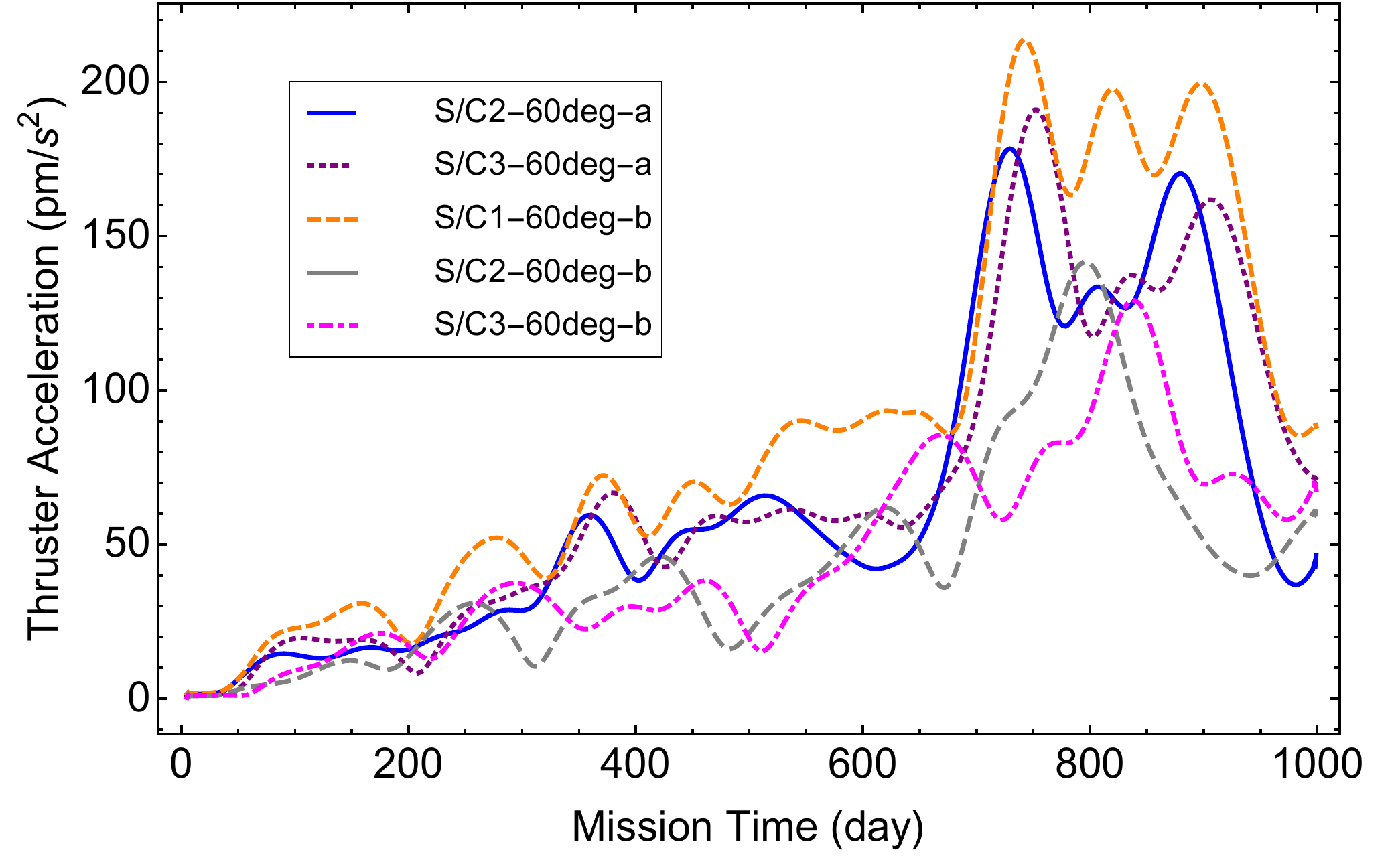} \ 
   \includegraphics[width=0.49\textwidth]{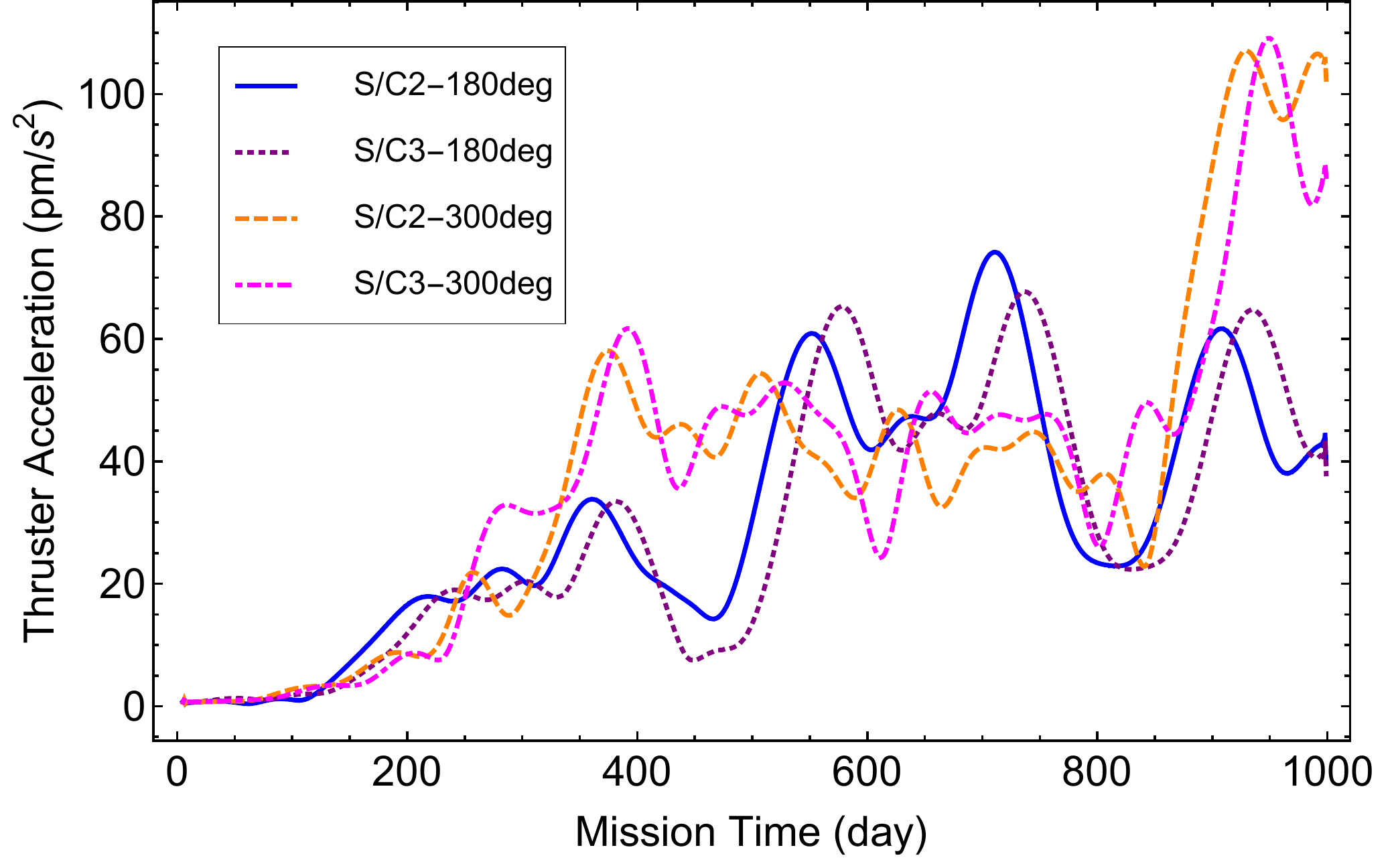}
   \caption{The thruster acceleration compensations to maintain the DECIGO arm length for S/Cs in the CGC3.0 ephemeris framework for 60-degree clusters a and b (left panel), 180-degree and 300-degree (right panel) trailing angle configurations, respectively.}
   \label{fig:diffA_DECIGO_2}
\end{figure}

We assume the scientific observation of DECIGO starts from $t_0 =$ JD2464694.0 (2036-Jan-1st 12:00:00) and calculation the mission orbits. In this work, we presume that two possible placements for the three constellations in the ecliptic plane which are
\begin{itemize}
\item three constellations at 30 deg (30-degree-a and 30-degree-b), 150 deg and 270 deg trailing angles with respect to position of the Earth, and the thruster acceleration compensations for each S/C at different constellations are shown in Fig. \ref{fig:diffA_DECIGO_1}.
\item three constellations at 60 deg (60-degree-a and 60-degree-b), 180 deg and 300 deg trailing angles with respect to position of the Earth, and the thruster acceleration compensations for each S/C at different constellations are shown in Fig. \ref{fig:diffA_DECIGO_2}.
\end{itemize}

As we can read from Fig. \ref{fig:diffA_DECIGO_1} and \ref{fig:diffA_DECIGO_2}, the thruster acceleration requirement is less than $\sim$200 pm/s$^2$ in 1000 days. In the left panel of Fig. \ref{fig:diffA_DECIGO_1} and \ref{fig:diffA_DECIGO_2}, the accelerations required for the five spacecraft change with the distance from the selected fiducial S/C1 in cluster a. An alternative way is to let one of the spacecraft of cluster b to follow a geodesic orbit and the other two spacecraft to follow this spacecraft; this may have the advantage that one spacecraft of cluster b will not be accelerated, and cluster a and cluster b will be more independent. Unlike other cases considered in this paper, we have not optimized the geodesic orbits given by the first selected orbit initial conditions; if optimization is implemented, the thruster requirement for DECIGO should be less than the result we estimated (although of the same order).

\subsection{New LISA and TAIJI} \label{subsec:NewLISA_TAIJI}

The LISA proposed to use the drag-free technology to leave S/Cs orbiting with the gravitational field and the arm length change with time. The new LISA plans to use $2.5 \times 10^6$ km nominal arm length\cite{LISA2017}. The TDI is required to achieve the equivalent near-equal interference paths. TAIJI mission proposed a LISA-like orbit with nominal arm length $3 \times 10^6$ km\cite{Taiji2017}.

\begin{figure}[htb]
   \centering
   \includegraphics[width=0.48\textwidth]{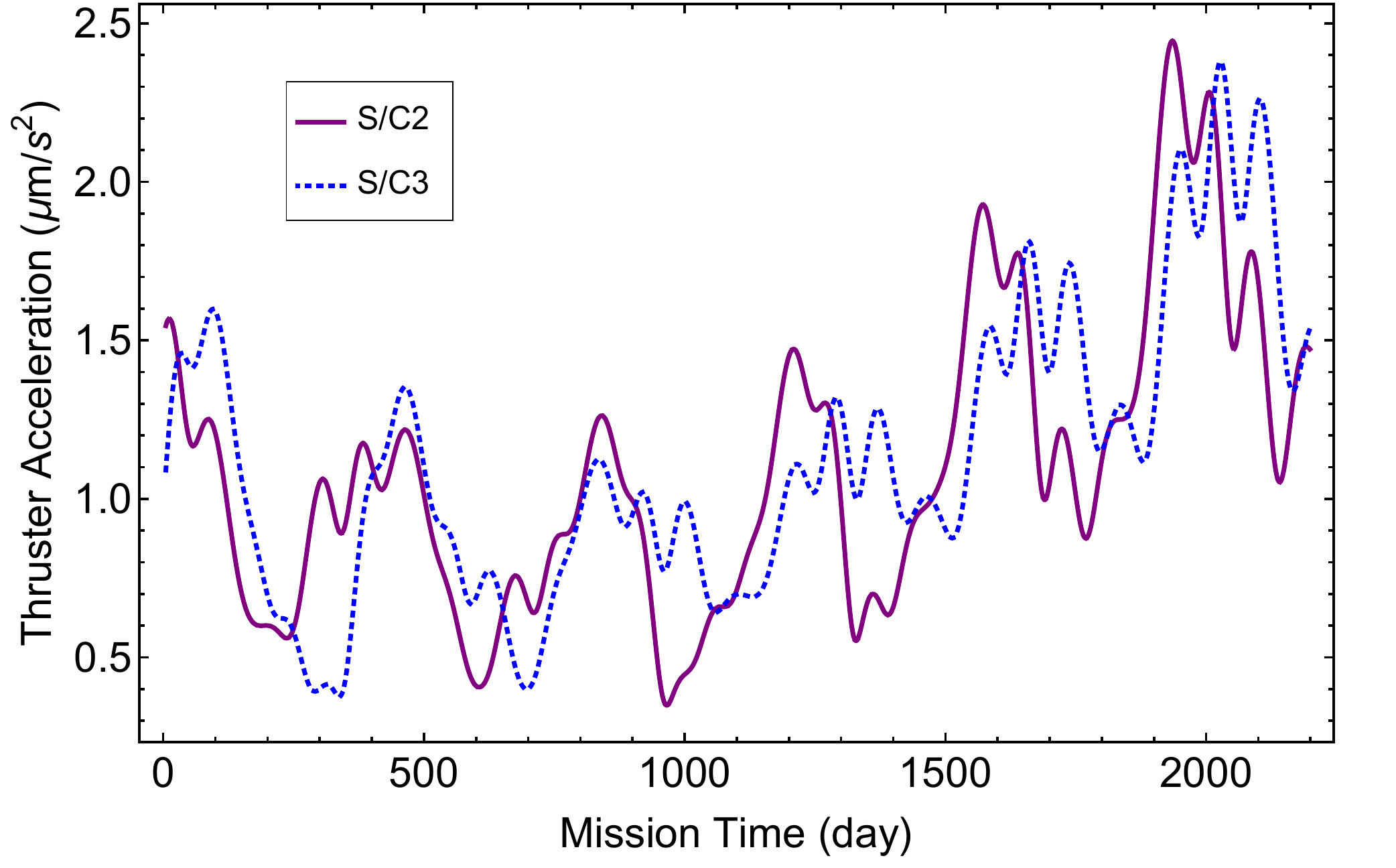} \ 
    \includegraphics[width=0.48\textwidth]{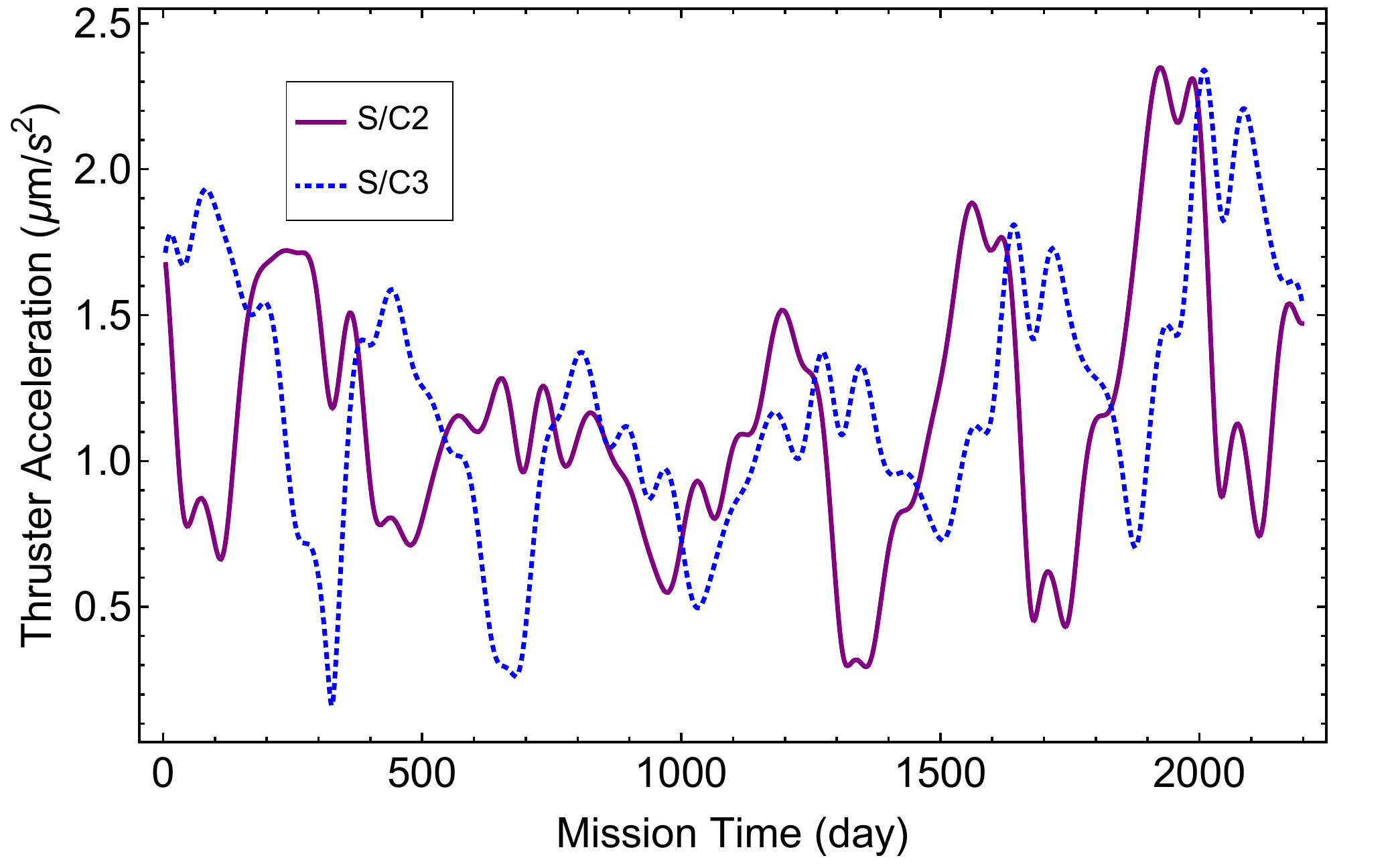} \
   \caption{The thruster acceleration compensations to maintain the New LISA constant arm length for a 20-degree trailing angle configurations (left panel), and the thruster acceleration compensation for the TAIJI for a 20-degree leading angle configuration (right panel).}
   \label{fig:diffA_LISA_TAIJI}
\end{figure}

In our previous work\cite{wang&ni2017}, we worked out a set of the LISA-like mission orbits with the observation starting time at $t_0 =$ JD2461853.0 (2028-Mar-22nd 12:00:00). Afterwards, we worked out another TAIJI mission orbit which is ahead of the Earth by 20$^\circ$ and form a $10^8$ km baseline with LISA. The large separation has high angular-resolution virtues for the short duration sources detection during joint LISA-TAIJI observation scenarios.
By using the orbits we obtained, we explore the thruster requirement for the LISA and TAIJI missions to maintain a constant arm configuration. From the method we described in Section \ref{sec:3}, the thruster acceleration compensations for the LISA and TAIJI mission are shown in Fig. \ref{fig:diffA_LISA_TAIJI}. The acceleration requirement could be up to 2.5 $\mu$m/s$^2$ for both LISA and TAIJI.

\section{Fuel Requirement at Different Specific Impulses} \label{sec:Fuel_result}

In this section, we assume that the mass $M$ of a typical spacecraft weighs 1000 kg and derive propellant requirement at two different specific impulses 300 sec and 1000 sec. Typical liquid fuels have about 300 sec specific impulse. The colloid thruster fuel has the specific impulse in the range of 200-2000 sec. Let $I_{\mathrm{sp}}$ be the specific impulse in unit of sec. The thruster force $F_\mathrm{thruster}$ of a propeller is
\begin{equation} \label{eq:F_thruster}
   F_{\mathrm{thruster}} (t) = g_0 I_\mathrm{sp} \dot{m} (t),
\end{equation}
where $g_0$ (= 9.8 m/s$^2$) is the gravity acceleration at surface of the Earth, and $\dot{m}$ (in unit of kg/s) is the rate of consumption of fuel. The consumed propellant $\Delta m$ is
\begin{equation} \label{eq:Delta_m_exp}
\Delta m = M \left[1 - \exp \left( - \frac{1}{g_0 I_{\mathrm{sp}} } \int_0^T a_\mathrm{thruster} dt \right) \right].
\end{equation}
For small $\Delta m / M$ ratio, it can be approximated as
\begin{equation} \label{eq:Delta_m}
\Delta m \approx \frac{1}{g_0 I_{\mathrm{sp}}} \int F_\mathrm{thruster} dt \simeq \frac{M}{g_0 I_{\mathrm{sp}}}  \int_0^T a_\mathrm{thruster} dt,
\end{equation}
where $T$ is the considered mission time. The yearly propellant in column 4 and 5 of Table \ref{Tab:results_table} is averaged by using
\begin{equation}
\overline{ \Delta m}_{\mathrm{yr}} = \Delta m \times \frac{1\ \mathrm{yr}}{T}
\end{equation}

From this formula, we calculate the propellant required for 1000 kg spacecraft in Table 1 at specific impulses 300 sec and 1000 sec of thruster assumed. We compile our results on the required acceleration, thruster requirement, and fuel requirement at different specific impulse 300 sec and 1000 sec for AMIGO, AIGSO, B-DECIGO, DECIGO, LISA, and TAIJI in Table \ref{Tab:results_table}. 
\begin{table}[ht]
\tbl{The thruster and propellant requirement for the different missions assuming the mass of the S/C is 1000 kg.}{
\centering 
\begin{tabular}{c|cc|cc}
 \hline \hline
Mission concept & Required & Thruster &  \multicolumn{2}{c}{Propellant requirement for 1 yr} \\
(arm length)  & acceleration  & requirement &  \multicolumn{2}{c}{by numerical integration (kg)} \\
 \cline{4-5}
 & (max) & (max) &  $I_{\mathrm{sp}}=300$ s  & $I_{\mathrm{sp}}=1000$ s  \\
\hline
AMIGO-E1 ($10^4$ km) & 2.0 mm/s$^2$ & 2.0 N & 999.8 & 922.0  \\
AMIGO-EML4 ($10^4$ km) & 2.5 mm/s$^2$ & 2.5 N & 863.0 & 449.2 \\
AMIGO-S ($10^4$ km) & 15 nm/s$^2$ & 15 $\mu$N & 5.0E-2 & 1.6E-2 \\
\hline
AIGSO-2$^\circ$-LISA-like (10 km) & 30 pm/s$^2$ & 30 nN & 1.1E-4 & 3.4E-5 \\
AIGSO-10$^\circ$-LISA-like (10 km) & 15 pm/s$^2$ & 15 nN & 2.8E-5 & 8.4E-6 \\
AIGSO-10$^\circ$-Ecliptic (10 km) & 10 pm/s$^2$ & 10 nN & 1.3E-5 & 4.0E-6 \\
\hline
B-DECIGO-AM-E1 (100 km) & 25 $\mu$m/s$^2$ & 25 mN & 97.7 & 30.4 \\
B-DECIGO-AM-EML4 (100 km) & 25 $\mu$m/s$^2$ & 25 mN & 19.7 & 6.0 \\
B-DECIGO-AM-S (100 km) & 0.15 nm/s$^2$ & 0.15 $\mu$N & 5.3E-4 & 1.6E-4 \\
\hline
DECIGO (1000 km) & 0.2 nm/s$^2$ & 0.2 $\mu$N & 8.0E-4 & 2.4E-4 \\
\hline
LISA ($2.5 \times 10^6$ km)	 & 2.5 $\mu$m/s$^2$ & 2.5 mN & 12.2 & 3.7 \\
\hline
TAIJI ($3 \times 10^6$ km) & 2.5 $\mu$m/s$^2$ & 2.5 mN & 13.0 & 3.9 \\
\hline \hline
\end{tabular} \label{Tab:results_table} }
\end{table}

As we can read from Eq. \eqref{eq:Delta_m}, the propellant requirement is proportional to the $M$ of the S/C and inversely proportional to the specific impulse $I_{\mathrm{sp}}$ if $\Delta m/M$ is small. In these cases, the propellant requirements could be readily inferred from Table \ref{Tab:results_table} for other values S/C mass and/or specific impulse. For the large $\Delta m/M$, the Eq. \eqref{eq:Delta_m_exp} should be applied.

\section{Discussions and Conclusions} \label{sec:conclusion}
For the geocentric orbit options of B-DECIGO which we have explored, the propellant mass requirement is a concern. For the heliocentric orbit options of B-DECIGO and DECIGO, the propellant requirement could be easily satisfied.

From propellant requirement estimated in Table \ref{Tab:results_table}, this scheme is technologically feasible for both LISA and TAIJI as far as fuel is concerned. This means that, for LISA and TAIJI, both ordinary Michelson interferometry and first-generation Michelson interferometry would be good configurations for GW detection if other accompanying requirements can be met.
The two accompanying requirements to fuel requirement of ordinary constant equal-arm Michelson interferometry implementation of LISA/TAIJI are (i) thruster noise requirement, (ii) and proof mass actuation requirement.

With large thruster power, the thruster noise increases. To minimize thruster noise, 2-stage thrusters/3-stage thrusters could be considered. With the pseudo-random code (PRC) ranging, sub-meter range accuracy can be achieved. A set of mN thrusters could provide the basic thruster force. The $\mu$N thruster could just provide the small residual precision adjustment to the acceleration needed to confirm to the PRC ranging measurement. The limiting noise would come from $\mu$N thruster noise and pseudo-random code range measurement noises.

The other requirement concerns proof mass actuation noise. In the application of $\mu$m/s$^2$ acceleration to proof mass, the noise should not be greater than pm/s$^2$ in some kind of average. A reference is needed for this measurement/monitoring of the actuation. Therefore an alternate proof mass is needed. Laser metrology has the required accuracy\cite{Luo}. The two proof masses can alternate to become the reference masses. This way the required dynamical range can be achieved. However, the gap size limits the total range of one acceleration maneuver to about 2 mm. This limits the one acceleration maneuver time of AMIGO-S to about 500 sec. For AIGSO, B-DECIGO-AM-S, and DECIGO, the time limits of one acceleration maneuver are about 5000 sec, 1500 sec and 1500 sec, respectively. For LISA and TAIJI, this time limit is about 15 sec. The actuation induced Fourier spectral components needs to be subtracted. Whether the time limits for various mission implementation are enough need to be studied experimentally\cite{Lei}.

Constant equal-arm Michelson interferometry is preferred if these technical issue can be resolved, because it does not have the complication of the TDI. Moreover, constant equal-arm Michelson interferometry and the TDI could both be tested at the beginning of science mission and worked out in the same mission if pre-mission preparation was done.

\section*{Acknowledgments}
We would like to thank Gerhard Heinzel (AEI), Hongbo Jin (NAOC), Jungang Lei (LSTT), Ziren Luo (IM, CAS), Liming Song(IHEP), Stefano Vitale (Trento U.), Bill Weber (Trento U.), Peng Xu (Lanzhou U.), and Zhen Yang (NSSC) for helpful discussions. This work was supported by National Key Research and Development Program of China under Grant Nos. 2016YFA0302002 and 2017YFC0601602, and Strategic Priority Research Program of the Chinese Academy of Sciences under grant No. XDB21010100.



\begin{thebibliography}{0}    

\bibitem{LVC2016} B. P. Abbott, et al., (LIGO Scientific Collaboration and Virgo Collaboration), {\it Phys. Rev. Lett.} {\bf 116} (2016) 061102, and references therein.
\bibitem{LVC2017} B. P. Abbott, et al., (LIGO Scientific Collaboration and Virgo Collaboration), {\it Phys. Rev. Lett.} {\bf 119} (2017)161101, and references therein.
\bibitem{LVC2018}
  B.~P.~Abbott {\it et al.} (LIGO Scientific Collaboration and Virgo Collaborations),
  arXiv:1811.12907 [astro-ph.HE].
\bibitem{Kuroda:2015owv}
  K.~Kuroda, W.-T.~Ni and W.-P.~Pan,
  {\it Int.\ J.\ Mod.\ Phys.\ D} {\bf 24} (2015) no.14,  1530031
\bibitem{Armano:2016bkm}
  M.~Armano, H. Audley and G. Auger {\it et al.},
  {\it Phys.\ Rev.\ Lett.}\  {\bf 116} (2016) no.23,  231101.
\bibitem{Armano:2018kix}
  M.~Armano, H. Audley and J. Baird {\it et al.},
  {\it Phys.\ Rev.\ Lett.}\  {\bf 120} (2018) no.6,  061101.
\bibitem{LISA2017}
  P. Amaro-Seoane, H.~Audley and S. Babak {\it et al.} (LISA Collaboration),
  arXiv:1702.00786 [astro-ph.IM].
\bibitem{Taiji2017}
  W.-R.~Hu and Y.-L.~Wu,
  {\it Natl.\ Sci.\ Rev.}\  {\bf 4} (2017) no.5,  685.
\bibitem{Luo2015} J. Luo  L.-S. Chen, H.-Z. Duan et al., {\it Class. Quantum Grav.} {\bf 33} (2016) 035010. 
\bibitem{wang&ni2017}
  G.~Wang and W.~T.~Ni,
  {\it Res.\ Astron.\ Astrophys.}\  {\bf 19} (2019) no.4,  058, arXiv:1707.09127.
\bibitem{Tianqin2019} B.-B. Ye, X. Zhang, M.-Y. Zhou et al., {\it Int. J. Mod. Phys. D} {\bf 28} (2019) 1950121.
\bibitem{Gao2017}
  D.~Gao, J.~Wang and M.~Zhan,
  {\it Commun.\ Theor.\ Phys.}\  {\bf 69} (2018) no.1,  37
\bibitem{Wang:2019oeu}
  G.~Wang, D.~Gao, W.-T.~Ni, J.~Wang and M.-S.~Zhan,
  {\it Int.\ J.\ Mod.\ Phys.\ D} {\bf 28} (2019) 1940004.
\bibitem{Ni:2017bzv}
  W.-T.~Ni,
  {\it EPJ Web Conf.}\  {\bf 168} (2018) 01004
\bibitem{AMIGO} W.-T. Ni, G. Wang and A.-M. Wu, Astrodynamical middle-frequency interferometric gravitational wave observatory AMIGO: Mission concept and orbit design, {\it Int.\ J.\ Mod.\ Phys.\ D} {\bf 28} (2019) 194000X.
\bibitem{ni2009} W.-T. Ni et al., ASTROD optimized for Gravitational Wave detection: ASTROD-GW (in Chinese), in Proceedings of Sixth Deep Space Exploration Technology Symposium, December 3-6, 2009, Sanya, Hainan, China, pp. 122-128 (2009).
\bibitem{ni2010} W.-T. Ni, {\it Mod. Phys. Lett. A} {\bf 25} (2010) 922.
\bibitem{ni2013} W.-T. Ni, {\it Int. J. Mod. Phys. D} {\bf 22} (2013) 1431004.
\bibitem{men2010a} J.-R. Men, W.-T. Ni, and G. Wang, {\it Chin.Astron.Astrophys.} {\bf 34} (2010) 434.
\bibitem{men2010b} J.-R. Men, W.-T. Ni, and G. Wang, {\it Acta Astronomica Sinica} {\bf 51} (2010) 198.
\bibitem{wang&ni2012} G. Wang and W.-T. Ni, {\it Chin.Astron.Astrophys.} {\bf 36} (2012) 211.
\bibitem{wang&ni2013CPB} G. Wang and W.-T. Ni, {\it Chin. Phys. B} {\bf 22} (2013) 049501.
\bibitem{wang&ni2015} G. Wang and W.-T. Ni, {\it Chin. Phys. B} {\bf 24} (2015) 059501.
\bibitem{Crowder&Cornish} J. Crowder and N. J. Cornish, {\it Phys. Rev. D} {\bf 72} (2005) 083005. 
\bibitem{B-DECIGO} S. Isoyama, H. Nakano and T. Nakamura, {\it Prog. Theor. Exp. Phys} (2018) 073E01.
\bibitem{Kawamura2018} S. Kawamura, T. Nakamura, M. Ando et al., {\it Int. J. Mod. Phys. D} {\bf 27} (2018) 1845001.
\bibitem{Kawamura2006} S. Kawamura et al., {\it Class. Quantum Grav.} {\bf 23} (2006) S125. 
\bibitem{Ni:2008bj}
  W.-T.~Ni,
  {\it Class.\ Quant.\ Grav.}\  {\bf 26} (2009) 075021
\bibitem{hogan2011} J. M. Hogan et al., {\it Gen. Relativ. Gravit.} {\bf 43} (2011) 1953.
\bibitem{hogan2016} J. M. Hogan and M. A. Kasevich, {\it Phys. Rev. A} {\bf 94} (2016) 033632.
\bibitem{graham2017} P. W. Graham, J. M. Hogan, M. A. Kasevich, S. Rajendran and R. W. Romani, Midband gravitational wave detection with precision atomic sensors, arXiv:1711.02225.
\bibitem{loeb2015} A. Loeb and D. Maoz, arXiv:1501.00996.
\bibitem{vutha2015} A. Vutha, {\it New J. Phys}. {\bf 17} (2015) 063030.
\bibitem{Kolkowitz2016} S. Kolkowitz et al., {\it Phys. Rev. D} {\bf 94} (2016) 124043.
\bibitem{Ebisuzaki:2018ujm}
  T.~Ebisuzaki, H.~Katori, J.~Makino, A.~Noda, H.~Shinkai and T.~Tamagawa,
  {\it Int. J. Mod. Phys. D} {\bf 28} (2019) 1940002, arXiv:1809.10317 [astro-ph.IM].
\bibitem{chaibi2016} W. Chaibi, R. Geiger, B. Canuel, A. Bertoldi, A. Landragin, and P. Bouyer, {\it Phys. Rev. D} {\bf 93} (2016) 021101(R).
\bibitem{canuel2018} B. Canuel \textit{et al.}, {\it Sci. Rep.} {\bf 8} (2018) 14064.

\bibitem{SOGRO} H. J. Paik, C. E. Griggs, M. V. Moody et al., {\it Class. Quantum Grav.} {\bf 33} (2016) 075003 
\bibitem{SOGRO2}
  H.~J.~Paik, M. V. Moody and R. S. Norton,
  {\it Int.\ J.\ Mod.\ Phys.\ D} {\bf 28} (2019), 1940001.
\bibitem{TOBA} A. Shoda, Y. Kuwahara, M. Ando, et al., {\it Phys. Rev. D} {\bf 95}, (2017) 082004. 
\bibitem{TOBA2} T. Shimoda, S.Takano, C.P. Ooi, N. Aritomi, A,Shoda, Y. Michimura, and M. Ando, {\it Int. J. Mod. Phys. D} {\bf 28} (2019) 1940003, arXiv:1812.01835.
\bibitem{Zhan:2019quq}
  M.-S.~Zhan, J. Wang, W.-T. Ni {\it et al.},
  {\it Int.\ J.\ Mod.\ Phys.\ D} {\bf 28} (2019) 1940005, arXiv:1903.09288 [physics.atom-ph].

\bibitem{dnw2013}  S. V. Dhurandhar, W.-T. Ni and G. Wang, {\it Adv. Space Res.} {\bf 51} (2013) 198.
\bibitem{wang&ni2013CQG} G. Wang and W.-T. Ni, {\it Class. Quantum Grav.} {\bf 30} (2013) 065011.
\bibitem{Dhurandhar+etal+2005} S. V. Dhurandhar, K. R. Nayak, S. Koshti and J. Y. Vinet, {\it Class. Quantum Grav.} {\bf{22}} (2005) 481-488.
\bibitem{wang2011} G. Wang, Time-delay Interferometry for ASTROD-GW (MS Thesis) (Nanjing: Purple Mountain Observatory, 2011) (in Chinese).
\bibitem{Luo} Z. Luo {\it et al}, Associated laser metrology study for various constant-arm GW implementation, in progress
\bibitem{Lei} J. Lei {\it et al}, Actuation noise estimation for various constant-arm GW implementation, in progress

\end{thebibliography}
\end{document}